\documentclass[preprint,preprintnumbers,nofootinbib,amsmath,amssymb]{revtex4-1}

\usepackage[english]{babel}
\usepackage{color}
\usepackage{amsmath}
\usepackage{physics}
\usepackage{xr-hyper}
\usepackage[hidelinks]{hyperref}
\usepackage{diagbox}
\usepackage{multirow}
\usepackage{amsfonts}
\usepackage{amssymb}
\usepackage{latexsym}
\usepackage{graphicx}
\usepackage{adjustbox}
\usepackage{float}
\usepackage{xr}
\usepackage{subcaption}
\usepackage{caption}
\usepackage{enumitem}
\usepackage{threeparttable}
\usepackage{ytableau}
\usepackage[nodisplayskipstretch]{setspace} 

\newcommand{\al}{\alpha}

\newcommand{\pa}{\partial}

\newcommand{\veps}{\varepsilon}

\newcommand{\de}{\delta}

\newcommand{\rar}{\rightarrow}
\newcommand{\lrar}{\leftrightarrow}

\newcommand{\non}{\nonumber}

\begin{document}

\title{Ultra-Compact accurate wave functions for He-like and Li-like iso-electronic sequences and variational calculus. II. Spin-singlet (excited) and spin-triplet (lowest) states of the Helium sequence}

\author{A.V.~Turbiner}
\email{turbiner@nucleares.unam.mx, alexander.turbiner@stonybrook.edu}

\author{J.C.~Lopez Vieyra}
\email{vieyra@nucleares.unam.mx}

\author{J.C.~del~Valle}
\email{delvalle@correo.nucleares.unam.mx}

\affiliation{Instituto de Ciencias Nucleares, Universidad Nacional Aut\'onoma de M\'exico,
A. Postal 70-543 C. P. 04510, Ciudad de M\'exico, M\'exico.}

\author{D.J.~Nader}
\email{daniel.nader@correo.nucleares.unam.mx}
\affiliation{ Facultad de F\'isica, Universidad Veracruzana, A. Postal 70-543 C. P. 91090, Xalapa, Veracruz, M\'exico. }


\begin{abstract}

As a continuation of Part I \cite{Part-1:2020} (Int. Journal of Quantum Chem. 2021; 121: qua.26586), dedicated to the ground state of He-like and Li-like isoelectronic sequences for nuclear charges $Z \leq 20$, a few ultra-compact wave functions in the form of generalized Hylleraas-Kinoshita functions are constructed, which describe the domain of applicability of the Quantum Mechanics of Coulomb Charges (QMCC) for the energies (4-5 significant digits (s.d.)) of two excited states of He-like ions: the spin-singlet (first) excited state $2^1 S$ and the lowest spin-triplet $1^3 S$ state. For both states it provides absolute accuracy for energy $\sim 10^{-3}$\,a.u., exact values for cusp parameters and also for 6 expectation values the relative accuracy $\sim 10^{-2}$. The Bressanini-Reynolds observation about the special form of the nodal surface of the $2^1 S$ state of Helium is confirmed and extended to He-like ions with $Z > 2$. Critical charges $Z=Z_B$, where ultra-compact trial functions lose their square-integrability, are estimated:
$Z_B(1^1 S)\approx Z_B(2^1 S)\sim 0.905$ and $Z_B(1^3 S)\sim 0.902$.
For both states the Majorana formula - the energy as a second degree polynomial in $Z$ - provides accurately 4-5 significant digits for $Z \leq 20$.
\end{abstract}

\maketitle
\newpage

\section*{Introduction}

Search for compact approximation of wave functions is one of interesting open directions in contemporary quantum mechanics. In the case of few-body Coulomb problems it allows to gain understanding of physics of interactions between bodies in small atoms and molecules \cite{Harris:2005,David:2006,TGH:2009,H3+} and in particular \cite{BMBM:2001}.
Likely, it was the original idea by E.A.~Hylleraas when he introduced the so-called ``Hylleraas function" for the helium atom \cite{Hylleraas:1929} (and performed concrete calculations) in two forms: with interelectronic distance $r_{12}$ in exponent, see its equation (18), as a precursor of what later was called the "Exponentially correlated Slater function“ by F.~Harris et al and the "Exponential Hylleraas function" by V.I.~Korobov et al, and in pre-factor, see for a discussion \cite{RSP:2021} and numerous references. It should be noted that many years after the Slater-Hylleraas pioneering works, J.O.~Hirschfelder \cite{Hirschfelder:1963} was the first who unified these two forms by writing the correlation $r_{12}$ in both exponent and pre-factor.

Contrary to the highly complicated variational trial functions composed by thousands of terms, which lead to highly accurate energies, compact wave functions can be easily interpreted. In particular, it is quite valuable if the wave function includes non-linear parameters, which admit a physical meaning  of charge screening, their optimal values provide understanding the physics picture of the Coulombic interactions in media. In this sense  Mulliken \cite{Mulliken:1965} expressed himself in favor of physical intuition or, saying differently, the qualitative physics picture behind contrary to the high accuracy. Needless to say that compact trial functions are specially valuable when they are sufficiently accurate locally (in coordinate space) to be able to reproduce the domain which is free of possible corrections of any type: relativistic, QED, finite mass, etc. For the He-like and Li-like isoelectronic sequences the correction-free domain for the ground state energy was localized in \cite{AOP:2019}: it was of order of 4 significant digits (s.d.). It is natural to assume this correction-free domain remains unchanged for excited states as well.

Perhaps, it should be also emphasized that the compact trial functions are not only valuable in order to gain understanding of quantum mechanics of compact few-body systems - bound state - but they are also of interest to the collision community \cite{Ancarani}. It turns out that simple but accurate wave functions are very useful as a starting point in the calculation of double ionization cross section by electron or radiation impact as was indicated as early as by Chandrasekhar at 1944 \cite{Chandrasekhar:1944}, as for recent references see e.g. \cite{Jones:2003,Bahati:2005} and references therein. Highly sophisticated wavefunctions with large numbers of terms and parameters are frequently non-practical since they require time-demanding computer codes to evaluate cross sections.

In Part I \cite{Part-1:2020} we introduced (ultra)-compact wave functions for He-like and Li-like iso-electronic sequences in their respective ground states with idea to get a description of the above-mentioned correction-less domain.
These ultra-compact wave functions with a few linear and non-linear variational parameters led
to accurate variational energies but also provide highly accurate expectation values while
satisfying cusp conditions with high accuracy.
Needless to say that in general it is not clear how accurate locally
the standard trial wave functions, in particular, made from terminated expansions,
are. They are certainly accurate in the domain giving the dominant contribution in the energy
integrals but could not be that accurate outside of the this domain. They can provide expectation values with much worse accuracy than energy, see for discussion the Part I, footnote 15 and also \cite{Chandrasekhar:1944}.

Note that the non-linear parameters of the ultra-compact trial wave functions can be systematically adjusted versus the nuclear charge $Z$ using the quadratic fit in $Z$ for $Z \leq 20$.
Moving in $Z$ the relative accuracy in energy is kept fixed: however, it even gets improved for larger nuclear charges. In general, for any of both He-like and Li-like sequences at $Z \leq 20$ the ground state energy is described by the second degree polynomial in $Z$ - the Majorana formula - with 4-5 significant digits, it corresponds to the correction-less region.

This paper is the second part of the series where we follow the same philosophy of design of
(ultra)-compact wave functions for the lowest excited $S$-states of para- and ortho-Helium,
and their respective isoelectronic sequences in the domain of nuclear charges $Z \leq 20$.
The structure of this paper is the following: in Section I we discuss the He-like,
two-electron sequence, considering the first excited spin-singlet state $2^1 S$ while
in Section II the lowest spin-triplet state $1^3 S$.
The results are summarized in Conclusions.

Atomic units are used throughout this paper.

\section{Spin-singlet excited state $2^1S$ (para-Helium sequence):
generalizing Hylleraas and Kinoshita functions}

\subsection{Generalities, exact solution}

In general, the orbital function for any spin-singlet state (para-Helium) with zero total angular momentum depends on relative distances only, see e.g. \cite{twe}. It is symmetric with respect to permutation of electrons,
\[
   \Psi (r_1,r_2,r_{12})\ =\ \Psi (r_2,r_1,r_{12})\ ,
\]
and usually can be represented as
\begin{equation}
\label{symm}
  \Psi (r_1,r_2,r_{12})\ =\ (1 + P_{12})\,\Phi (r_1,r_2,r_{12})\ =\ \Phi (r_1,r_2,r_{12}) +
  \Phi (r_1,r_2,r_{12})\ ,
\end{equation}
where $P_{12}$ is permutation operator $(1 \lrar 2)$.

The orbital function obeys to the reduced Schr\"odinger equation
\begin{equation}
\label{SE-radial}
  \hat{H}\,\Psi (r_1,r_2,r_{12})\ =\ E\,\Psi (r_1,r_2,r_{12})\ ,
\end{equation}
where $\hat{H}=\hat{H}(r_1,r_2,r_{12})$,
\begin{gather}
   \hat{H} \ =\ -\frac{1}{2}
\left[
  \frac{\pa^2}{\pa r_1^2}\ +\ \frac{\pa^2}{\pa r_2^2}\ +\ 2\frac{\pa^2}{\pa r_{12}^2}\ +\
         \frac{2}{r_1}\frac{\pa}{\pa r_1}\ +\ \frac{2}{r_2}\frac{\pa}{\pa r_2}\ +\
                     \frac{4}{r_{12}}\frac{\pa}{\pa r_{12}}
\right. \non
\\
\left.
  +\ \left(
  \frac{r_1^2-r_2^2+r_{12}^2}{r_1r_{12}}\right)\frac{\pa^2}{\pa r_1 \pa r_{12}}
  \ +\
  \left(\frac{r_2^2-r_1^2+r_{12}^2}{r_2r_{12}}\right)\frac{\pa^2}{\pa r_2 \pa r_{12}}
    \right]
\non\\
  +\ \left[\frac{-Z}{r_1}\ +\ \frac{-Z}{r_2}\ +\ \frac{1}{r_{12}}\right]\ ,
\label{Hred}
\end{gather}
see in \cite{GAM:1987} the Eq.(5), is the so-called {\it radial} 3-body Schr\"odinger Hamiltonian \cite{twe}.
Here, $r_1$ and $r_2$ are the electron distances from the nucleus of charge $Z$, while
$r_{12}$ is interelectronic distance.

From (\ref{Hred}) it is easy to obtain the first terms of the so-called Fock expansion for orbital function
of arbitrary spin-singlet, permutationally-symmetric $(r_1 \lrar r_2)$ state
\begin{equation}
\label{fock-2s}
\Psi\ =\ 1\ -\  C_{Z,e}\,(r_1+r_2)\ +\ C_{e,e}\,r_{12}\ +\ \ldots\ ,
\end{equation}
where the coefficients
\begin{equation}
C_{Z,e}\  =\ Z\quad ,\quad C_{e,e}\ =\ \frac{1}{2}\ ,
\end{equation}
are called the \textit{cusp} parameters. They have a meaning of residues in Coulomb singularities of the potential.

Making in (\ref{SE-radial})-(\ref{Hred}) the scale transformation,
\[
   r \rar u = Z r\ ,
\]
one can see that the kinetic energy operator in $\hat{H}$ remains unchanged up to the multiplicative factor $Z^2$, while the potential energy and the spectral parameter $E$ are changed. We arrive at the radial Schr\"odinger like equation
\begin{equation}
\label{SE-radial-u}
  \hat{H}\,\Psi (u_1,u_2,u_{12})\ =\ \veps\,\Psi (u_1,u_2,u_{12})\ ,
\end{equation}
where
\begin{gather}
   \hat{H} \ =\ -\frac{1}{2}
\left[
  \frac{\pa^2}{\pa u_1^2}\ +\ \frac{\pa^2}{\pa u_2^2}\ +\ 2\frac{\pa^2}{\pa u_{12}^2}\ +\
         \frac{2}{u_1}\frac{\pa}{\pa u_1}\ +\ \frac{2}{u_2}\frac{\pa}{\pa u_2}\ +\
                     \frac{4}{u_{12}}\frac{\pa}{\pa u_{12}}
\right. \non
\\
\left.
  +\ \left(
  \frac{u_1^2-u_2^2+u_{12}^2}{u_1 u_{12}}\right)\frac{\pa^2}{\pa u_1 \pa u_{12}}
  \ +\
  \left(\frac{u_2^2-u_1^2+u_{12}^2}{u_2 u_{12}}\right)\frac{\pa^2}{\pa u_2 \pa u_{12}}
    \right]
\non\\
  +\ \left[-\frac{1}{u_1}\ -\ \frac{1}{u_2}\ +\ \frac{1}{Z}\ \frac{1}{ u_{12}}\right]\ ,\quad
  \veps=\frac{E}{Z^2} \ ,
\label{Hred-u}
\end{gather}
plays the role of the Hamiltonian, it describes two hydrogen atoms with interelectron repulsion and unusually-written kinetic energy.


At $Z \rar \infty$ the spectral problem (\ref{SE-radial-u}) with the Hamiltonian (\ref{Hred-u}) corresponds to two non-interacting hydrogen atoms, $u_{12}$ dependence disappears and variables $u_1, u_2$ are separated. As a realization of permutation symmetry $(1 \lrar 2)$ the exact eigenfunctions appear as the sum of symmetrized products of two Coulomb orbitals. The ground state $(1s1s\,1 {}^1 S)$ can be presented as
\begin{equation}
\label{He-infty-psi-0}
  \Psi_0\ =\ \frac{1}{2}(1 + P_{12})\,e^{-\al u_1 - \beta u_2}\ \sim
  (1s_1 1s_2) + (1s_2 1s_1) \ ,
\end{equation}
when $\al=\beta=1$, the ground state energy $E_0=-Z^2$ is twice of the ground state of the hydrogen atom. In turn, the first excited state $(1s2s \, 2 {}^1 S)$ is made from symmetrized $(2s_1 1s_2)$ hydrogenic Coulomb orbitals,
\begin{equation}
\label{He-infty-psi-1}
  \Psi_1\ =\ \frac{1}{2}(1 + P_{12})\,(1 - a u_1)\,e^{-\al u_1 - \beta u_2}\
  \sim (2s_1 1s_2) + (1s_1 2s_2)\ ,
\end{equation}
where $a=1/2, \al=1/2, \beta=1$ with eigenvalue
\begin{equation}
\label{He-infty-E-1}
  E_1\ =\ -  \frac{5 Z^2}{8} \ ,
\end{equation}
which is evidently orthogonal to the ground state (\ref{He-infty-psi-0}).
The nodal surface, where $\Psi_1=0$, is symmetric $u_1 \lrar u_2$, it starts at the point $u_1=u_2=2$ and then it goes to the end-point $u_1(u_2)=0$, the value of $u_2(u_1)=2.556929$.

Needless to say that the function $\Psi_1$ (\ref{He-infty-psi-1}) can be used as trial function for any integer $Z=2,3,\ldots$ if the orthogonality condition to the ground state, taken for instance in the form (\ref{He-infty-psi-0}), is imposed. It fixes the parameter $a$ in terms of $\al,\beta$, those eventually are used as variational parameters. It resembles the
function used by Stillinger and Stillinger for the ground state $(1s1s\,1 {}^1 S)$ of He-like sequence \cite{Stillinger:1974}.

\subsection{Compact trial functions}

Natural generalization of (\ref{He-infty-psi-1}), where interelectron distance is involved explicitly, emerging from interpolation between Hylleraas-exponential-type $(b=0)$ \cite{Hylleraas:1929} and Kinoshita-type $(a=0, \gamma=0)$ \cite{Kinoshita} functions
\footnote{ In fact, the Kinoshita-type function was proposed for the first time by S.~Chandrasekhar for the case of the ground state, see \cite{Chandrasekhar:1944}, Eq.(7). However, when one of the authors (AVT), being a student, took the class on the advanced quantum mechanics \cite{AVT:1972} this function was called {\it the Kinoshita function} as a simplification of the functions presented in \cite{Kinoshita} (we followed this name in our Part I \cite{Part-1:2020}) and keeping in minds the fact that T.~Kinoshita was the first who indicated this function can be used for the excited states, in particular, for those we study in present paper.} has the form
\begin{equation}
\label{He-infty-psi-2}
  \Psi_{HK}\ =\ \frac{1}{2}(1 + P_{12}) (1 - a Z r_1 + b r_{12})\,e^{-\al Z r_1 - \beta Z r_2  + \gamma r_{12}}\ ,
\end{equation}
cf.(\ref{symm}), which can be further generalized to
\begin{equation}
\label{He-infty-psi-3}
  \Psi_{G}\ =\ \frac{1}{2}(1 + P_{12}) (1 - a Z r_1 + b r_{12})\,e^{-\al Z {\hat r}_1 - \beta Z r_2  + \gamma {\hat r}_{12}}\ .
\end{equation}
In the latter function the effects of screening, which are different for small and large distances, are taken into account by replacing $r_1 \rar {\hat r}_1$ and $r_{12} \rar {\hat r}_{12}$ in the exponential, see below. From the physics viewpoint the first terms in exponential of $\Psi_{HK}, \Psi_G$ reflect the fact that the second electron is situated (in average) closer to the nucleus than the first one, $\al > \beta$ (it is the so-called clusterization effect). Hence, the interaction of the 2nd electron with nucleus can be considered as non-screened, $\beta \sim 1$, unlike the 1st one, for which the interaction with nucleus is screened by the presence of the second electron. Therefore, for the 1st electron the screening should be taken into account explicitly, for example, like
\[
   \al r_1 \rar \al {\hat r}_1 \equiv \al r_1 \frac{1 + c r_1}{1 + d r_1}\ ,
\]
see \cite{BMBM:2001}.
In a similar way the screening of the Coulomb repulsion between electrons due to
\pagebreak
presence of nucleus somehow in between of them should be turned on, for example, in the form,
\[
   \gamma r_{12} \rar \gamma {\hat r}_{12} \equiv \gamma r_{12} \frac{1 + c_{12} r_{12}}{1 + d_{12} r_{12}} \ .
\]
It depends on interelectronic distance.
In order to perform concrete calculations for the excited state $(1s2s \, 2 {}^1 S)$ the orthogonality condition $(\Psi^{(ground\ state)}, \Psi_1)=0$ to the ground state should be imposed. It allows to fix one of the parameters (we choose the parameter $a$) and eventually the trial functions (\ref{He-infty-psi-2}), (\ref{He-infty-psi-3}) become 4- and 8-parametric, respectively.
The ground state function was taken from Part I \cite{Part-1:2020} in two forms, as the Hylleraas-Kinoshita function $\Psi^{(ground)}_{HK}$, see Eq.(17) in Part I, and the seven-parametric generalized Hylleraas-Kinoshita function $\Psi^{(ground)}_{F}$, see Eq.(23) in Part I, with fitted parameters Eq.(28). Then the variational calculation was carried out. The results of calculations are presented in Table~\ref{Table1} for $Z=3/2,\ 2,\ 10,\ 20$. Surprising observation is that the accurate variational energies are compatible with $b=0$ for both trial functions (\ref{He-infty-psi-2}), (\ref{He-infty-psi-3}), independently on $Z$, of course, if the accuracy of 4-5 s.d. is considered. It reduces effectively a number of variational parameters. Inside of this accuracy we did not observe the dependence on the variational results on the ground state function chosen in Part I \cite{Part-1:2020}. Essentially, the parameter $(Za)$ depends on $Z$ linearly, see below.

\begin{table}[h]

\begin{center}
	\caption{\label{Table1}Variational energy of first spin-singlet excited state $E_1$ of
    the helium-like atom at $Z=3/2,2,10,20$ calculated with $\Psi_{HK}$ (\ref{He-infty-psi-2})
    and $\Psi_{G}$ (\ref{He-infty-psi-3}). Orthogonality constraint imposed with respect to the Hylleraas-Kinoshita function for the ground state, see \cite{Part-1:2020}, Eq.(17) 
    and $\Psi_F$, see \cite{Part-1:2020} Eq.(23) with parameters (28), respectively. Electron-nuclear cusp $C_{Z,e}$ and electron-electron cusp $C_{e,e}$ shown (ratio of expectation values is on first row, while second row results found via the expansion  (\ref{fock-2s}) of $\Psi_{HK}$),\\
    on the third row the cusp parameters fixed to be exact and the results of constrained minimization of the variational energy for $\Psi_{G}$ (\ref{He-infty-psi-3}) shown.}
		{\setlength{\tabcolsep}{0.5cm}
			\begin{tabular}{c|cccccc}
				\hline			
				$Z$ & $E^{HK}$  & $E^{G}$    & $C_{Z,e}$ & $C_{e,e}$ & $E_{ref}$                   & $E^{(2S)}_M$ \\
				\hline
				\rule{0pt}{4ex}		
				3/2 & -1.1654   &            & 1.4963    & 0.0199    & -                           & -1.16755 \\
				    &           &            & 1.1353    & 0.0199    &                             &          \\
                                    &           & -1.1663    & 1.5       & 0.5       &                             &          \\[5pt]
				2   & -2.1438   &            & 1.9864    & 0.0363    & -2.1460$^{\dagger}$          & -2.14552 \\
				    &           &            & 1.6485    & 0.0363    & -2.1460$^{\star\star}$        &          \\
                                    &           & -2.1460    & 2.0       & 0.5       &                             &          \\[5pt]
				10  & -60.2882  &            & 9.9322    & 0.0956    & -60.2953$^{\star}$           & -60.2931 \\
				    &           &            & 9.6886    & 0.0956    & -60.2951$^{\star\star}$       &          \\
                                    &           & -60.2946   & 10.0      & 0.5       & -60.2953$^{\dagger\dagger}$   &          \\[5pt]
				20  & -245.4709 &            & 19.9265   & 0.1032    & -245.4776$^{\star}$          & -245.478 \\
				    &           &            & 19.7000   & 0.1032    & -245.4769$^{\star\star}$      &          \\
                                    &           &  -245.4770 & 20.0      & 0.5       &                             &          \\[5pt]
\hline
		\end{tabular}}
		\begin{tablenotes}
			\small
			\item $^{\dagger}$ non-rounded results: $E = -2.145\,974\,021$ \cite{Liverts},\   $E= -2.145\,974\,046\,054$\ \cite{Bressanini:2005-7},\\ $E = -2.145\,974\,046\,054\,419(6)$ \cite{Drake:2006}, $E = - 2.145\,974\,046\,054\,417\,415\,799 $ \cite{Pachucki:2010}; all four coincide in 8 s.d.
            \item $^{\star}$ calculated using the code provided in \cite{Liverts}, $^{\star\star}$ rounded results from ref. \cite{Drake:1988}, $^{\dagger\dagger}$ result from ref. \cite{Pachucki:2010}
		\end{tablenotes}
	\end{center}
\end{table}
\begin{table}[htp]
\caption{\label{2s1EVs}
   Expectation values for the excited state $(1s2s\,2{}^1S)$ using the function $\Psi_G$, constrained
   (first row) and unconstrained (second row).
   Results by Accad-Pekeris-Schiff  \cite{Accad:1971}, by Braun-Schweizer-Herold \cite{Braun:1993} and Liverts-Barnea \cite{Liverts} ${}^{(\star)}$ included for comparison. Electron-nuclear cusp $C_{Z,e}$ and electron-electron cusp $C_{e,e}$ found via the expansion  (\ref{fock-2s}) of $\Psi_{G}$, i.e.  $C_{Z,e} = ((\al+\beta) Z + a )/2$, $C_{e,e} = b +\gamma$. }
\begin{center}
{\setlength{\tabcolsep}{0.1cm} \renewcommand{\arraystretch}{0.5}
 \resizebox{0.9\textwidth}{!}{%
\begin{tabular}{|c|lllccccccc|}\hline
 $Z$																	&  $E$				& $C_{Z,e} $		  	&  $C_{e,e}$	& $\langle \delta({\mathbf r_1})\rangle$	& $\langle \delta({\mathbf r_{12}})\rangle$
															& $\langle r_{1}\rangle$	 & $\langle  r_{1}^2 \rangle$	& $\langle  r_{12} \rangle$
															& $\langle r_{12}^2 \rangle$ & source\\
\hline
       & 	& 	& 	&         &          &   &    &    &   &   \\[3pt]
3/2 & -1.16626  	& 1.5	& 0.5	& 0.5432 	     & 0.00154 	& 5.16634 & 51.67048 & 9.40825  & 103.51640 &   \\[3pt]
	& 	                  	& 1.59279 & 0.58884 & 0.5439	        & 0.00135        & 5.09344  & 49.77424 & 9.26241  & 	99.72620    &  \\[15pt]
2	& -2.14600    	& 2.0	& 0.5	& 1.3195	     & 0.01358  & 2.88920 & 15.34402 & 5.10698  & 30.7994  & \\[3pt]
	& 	                  	& 2.05887 & 0.53947 & 1.3079  & 0.00887 & 2.97208  & 16.08416 & 5.26699 & 	 32.27204 &
\\[3pt]
	& 	            & 	    & 	      & 1.3095  & 0.00865 & 2.97306	& 16.0891  & 5.26969  & 32.302  & \cite{Accad:1971}
\\[3pt]
	& 	            & 	    & 	      & 1.3094  & 0.00866 & 2.97321 &          & 5.26959  &         & 	\cite{Liverts}\\[3pt]
	& 	           	& 	    & 	      &         &         & 2.97318	&   &    & 32.18126	        &      \cite{Braun:1993}\\[15pt]
10	& -60.29456  	& 10.0	& 0.5	& 175.1837	& 5.87751 & 0.40185  & 0.26477  & 0.67128 & 0.53092  &
\\[3pt]
	& 	            & 9.94772 & 0.41940 & 174.6227 & 5.82404 & 0.40297 & 0.26626 & 0.67341 & 0.53392 &
\\[3pt]
    & 	            & 	      & 	    & 174.8737 & 5.74    & 0.40289 & 0.26584 & 0.67337 & 0.53338 & \cite{Accad:1971}
\\[5pt]
	&               & 	      & 	    & 174.8733 & 5.72932 & 0.40289 &         & 0.67337 &         & 	\cite{Liverts}
\\[15pt]	
20	& -245.47700  	& 20.0 	 & 0.5      & 1417.3854	& 54.86331  & 0.19381 & 0.06084 & 0.32131 & 0.12186 & \\[3pt]
	& 	           	& 19.87555 & 0.40990 & 1412.6008 & 54.29124 & 0.19413 & 0.06103 & 0.32190 & 0.12224 & \\[5pt]
	& 	            & 	       & 	     & 1415.0882 & 53.88928 & 0.19416 &         & 0.32202 &         & 	\cite{Liverts}\\[10pt]					

\hline
\end{tabular}}}
		\begin{tablenotes}
			\small
            \item $^{\star}$ calculated using the code provided in \cite{Liverts}
		\end{tablenotes}
\end{center}
\end{table}%

In general, the variational parameters in $\Psi_{G}$  (\ref{He-infty-psi-3}) leading to the minimal energy for the spin-singlet excited state $E_1$ of  the helium-like atom at $Z=3/2,2,10,20$  have a smooth behavior as a function of the nuclear charge $Z$. In particular, the parameters of the constrained minimization which is required to reproduce exact cusp parameters $C_{Z,e}= Z$ and $C_{e,e}= 1/2$, can be fitted for $2 \leq Z \leq 20$,
\begin{align}
Z\,a  &= -0.297218  + 0.505737 Z - 0.0001968 Z^2   \ ,\non  \\
b      &=0 \ ,\non \\
c_{12} &=  0.0405438  - 0.032151  Z + 0.00031 Z^2 \ ,\non \\
d_{12} &= 0.24351 + 0.111263 Z + 0.001843 Z^2 \ ,\non \\
\alpha &= 0.358158/(Z + 0.2528) + 0.489224 + 0.000358 (Z + 0.2528)  \ , \\
\beta  &=   2 - a - \alpha   \ ,\non \\
\gamma &= 1/2  \ ,\non \\
c   &=  -0.146884 + 0.162799 Z + 0.000512 Z^2 \ ,\non \\
d  & =  0.123749 + 0.164107 Z + 0.0005 Z^2\, .\non
 \end{align}
The parameter $a$, given by fit, guarantees approximate orthogonality of $\Psi_G$ to ground state $\Psi_F$ defined in Part I \cite{Part-1:2020}.
With above fitted parameters the expectation values for energy $E$ which coincide with those presented in
Table~\ref{Table1}.
It can be shown that the Majorana formula holds for this state
\begin{equation}
\label{Majorana2S-}
E^{(2^1S)}_M\ =\  -\frac{5 Z^2}{8}\ +\ 0.231547 Z\ -\ 0.108618\ .
\end{equation}
It provides accuracy 4 s.d. for $2 \leq Z \leq 20$, see Table I, in a similar way as for the ground state \cite{Part-1:2020}.

Six expectation values for the state $(1s2s\,2{}^1S)$ calculated using the function $\Psi_G$
with unconstrained (free) variational parameters and with variational parameters constrained
to reproduce exact cusp values are shown in Table \ref{2s1EVs}. There exists an agreement with results
by Accad-Pekeris-Schiff \cite{Accad:1971}, Braun-Schweizer-Herold \cite{Braun:1993} and \cite{Liverts}, wherever possible, in 2-3-4 s.d.

\subsection{Nodal surface}

Separate issue of the study is related with form of the nodal surface, where $\Psi_1(r_1,r_2,r_{12})=0$, for the state $(1s2s \, 2 {}^1 S)$. This nodal surface divides the coordinate space
into two subspaces. Usually, the localization of nodal surface is a difficult task, since in its vicinity the wavefunction is small, which can be smaller than accuracy of any approximate method used. It is worth noting that accurate localization of the nodal surfaces is crucial for employing Monte-Carlo technique for the excited states.

In \cite{Bressanini:2005-7} it was conjectured that for $Z=2$ the nodal surface for the state $(1s2s \, 2 {}^1 S)$ has a very weak dependence on the angle between vectors ${\bf r}_{1,2}$. At $Z=\infty$ the known exact solution (\ref{He-infty-psi-1}) confirms trivially its validity. Taking functions $\Psi_{HK}, \Psi_G$ the conjecture is checked for several $Z \in [2, 20]$, Figs.~1~-~3.
One can see that for both trial functions the nodal line, which lies in the first quadrant of $(u_1,u_2)$ plane, in fact, it has almost no dependence on angle ${\widehat {({\bf r_1},{\bf r_2})}}$ inside of the accuracy of variational method we used, it is defined by the thickness (!) of the drawing lines presented in Figs.~1~-~3. Nodal lines for different $Z$ look similar with rather weak dependence on $Z$. Hence, the conjecture by Bressanini and Reynolds \cite{Bressanini:2005-7} holds for nuclear charges $Z$ other than $Z=2$.
\begin{figure}
\caption{
\label{Z2}
Nodal line for the state $(1s2s \, 2 {}^1 S)$ for $Z=\infty$ and for $Z=2$ for $\Psi_{HK}$ (\ref{He-infty-psi-2}), $\Psi_{G}$ (\ref{He-infty-psi-3}) and comparison with nodal line, marked by $\Psi_S$, found in \cite{Bressanini:2005-7}.
}
  \includegraphics[scale=0.5,angle=0]{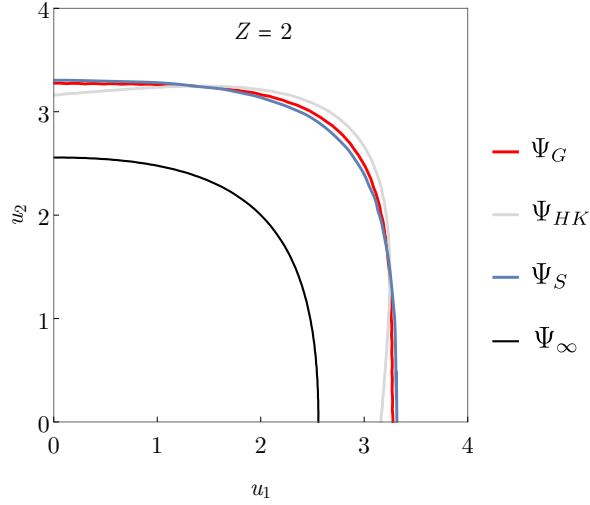}
\end{figure}
\begin{figure}
\caption{
\label{Z10}
Nodal line for the state $(1s2s \, 2 {}^1 S)$ for $Z=\infty$ and for $Z=10$ for $\Psi_{HK}$ (\ref{He-infty-psi-2}), $\Psi_{G}$ (\ref{He-infty-psi-3}).
}
  \includegraphics[scale=0.5,angle=0]{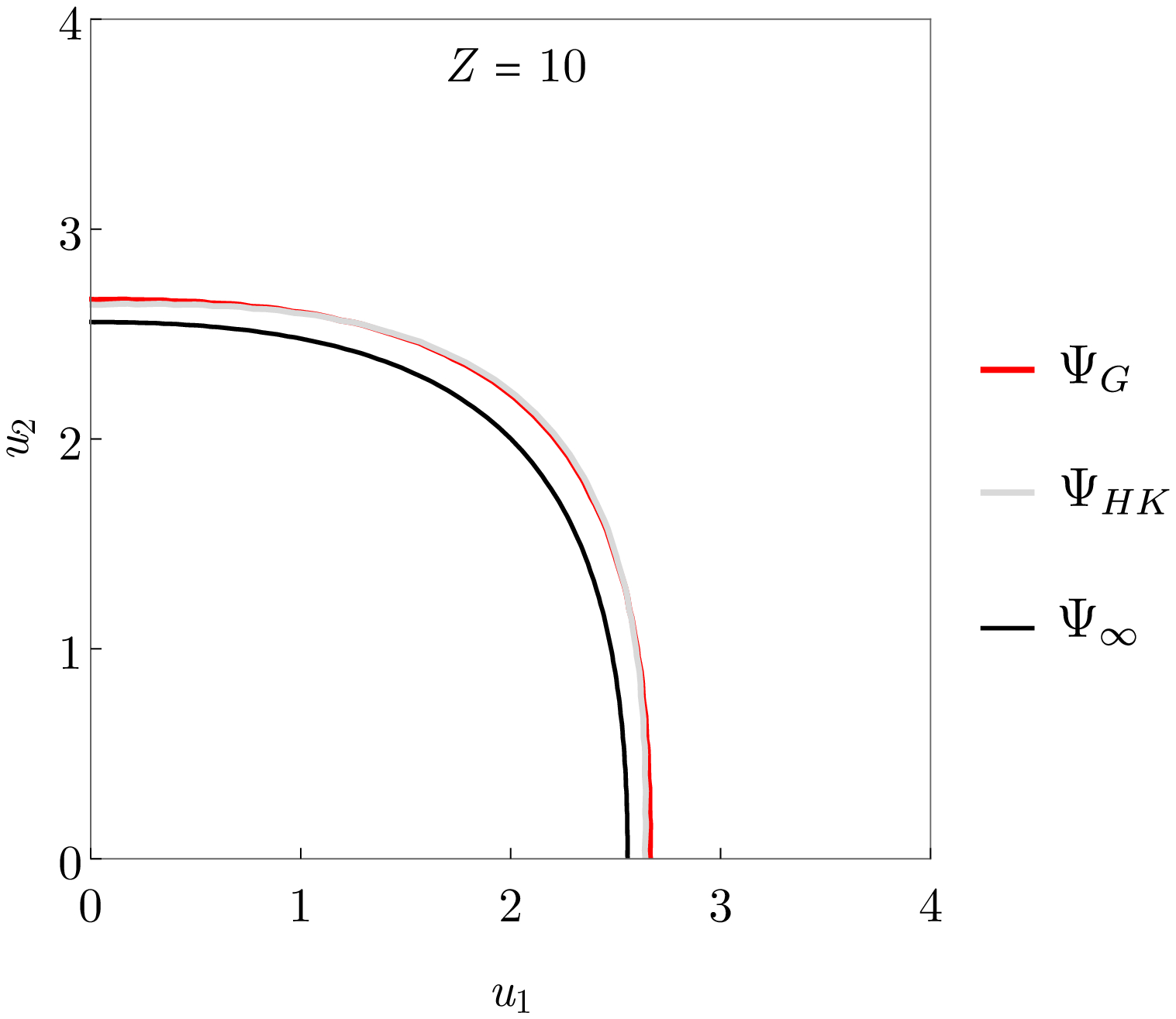}
\end{figure}
\begin{figure}
\caption{
\label{Z20}
Nodal line for the state $(1s2s \, 2 {}^1 S)$ for $Z=\infty$ and for $Z=20$ for $\Psi_{HK}$ (\ref{He-infty-psi-2}), $\Psi_{G}$ (\ref{He-infty-psi-3}).
}
  \includegraphics[scale=0.5,angle=0]{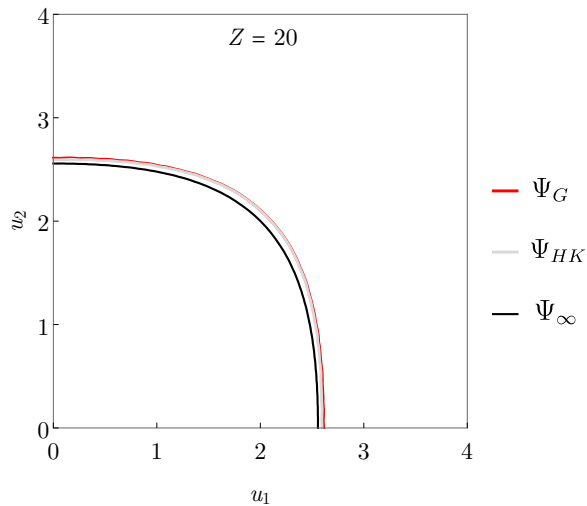}
\end{figure}

\newpage
\subsection{Square-integrability of compact functions}

{
It is evident that for $Z > 1$ keeping the position of one electron fixed the wavefunction decays exponentially at large values of the position of another electron, see for illustration Fig.\ref{large_distance}. It corresponds to the interaction of hydrogen-like ion $(Z e)$ with electron.
\begin{figure}
\caption{
\label{large_distance}
Configuration which corresponds to the interaction of hydrogen-like ion $(Z e)$ with electron.
}
  \includegraphics[scale=0.5,angle=0]{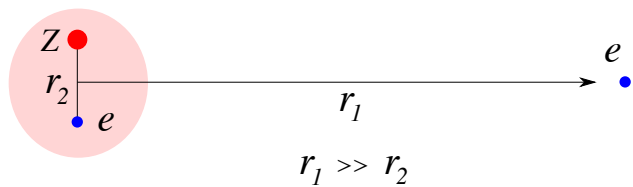}
\end{figure}
It can be explicitly seen if we take one of terms in the representation (\ref{symm}) for the wavefunction,
\begin{equation}
\label{phirinfinity}
 \Phi_1|_{r_2(r_1) fixed} \rar \exp {-A_{1,2}\, r_1(r_2)}\ \mbox{at}\ r_1(r_2) \rar \infty \ .
\end{equation}
In particular, at $Z \rar \infty$ for the exact ground state function (\ref{He-infty-psi-0}) in the case of the first term it gives
\begin{equation}
\label{GS-rates-exact}
     A^{(0)}_1\ =\  Z\ ,\ A^{(0)}_2\ =\  Z\ ,
\end{equation}
while for the first excited state $(1s 2s)$ the exact wavefunction (\ref{He-infty-psi-1}) leads to,
\begin{equation}
\label{ES-rates-exact}
     A^{(1)}_1\ =\ \frac{1}{2} \, Z\ ,\ A^{(1)}_2\ =\ Z\ .
\end{equation}
Straightforwardly, it implies that at $Z=Z^{(0,1)}_B=0$ square-integrability of (\ref{He-infty-psi-0}),(\ref{He-infty-psi-1}) is lost, which is well beyond of the domain where the exact solutions make sense.

Let us take the first term of $\Psi_G(r_1,r_2, r_{12})$ (\ref{He-infty-psi-3}) for the excited state $(1s2s \, 2 {}^1 S)$, which is the most accurate compact wave function we constructed. It is easy to find that
\begin{equation}
\label{AES}
     A^{ES}_1\ =\ \alpha Z \frac{c}{d} - \gamma \frac{c_{12}}{d_{12}}\ ,\
     A^{ES}_2\ =\ \beta Z  - \gamma \frac{c_{12}}{d_{12}}\ .
\end{equation}
Similar formulas occurs for the second term.
In concrete calculations one can find that for all studied $Z \leq 20$
\[
    A^{ES}_1\ < \ A^{ES}_2\ ,
\]
it signals the appearance of clusterization effect: the electron 2 is closer to the nuclei than the electron 1. The rate of convergence of variational integrals is defined by $A^{ES}_1$.

In a similar way the analysis can be performed for the ground state, see Part I, Eq.(23), which leads to
\begin{equation}
\label{AGS}
   A^{GS}_1\ =\ \alpha Z - \gamma \frac{c_{12}}{d_{12}}\ ,\
   A^{GS}_2\ =\ \beta Z  - \gamma \frac{c_{12}}{d_{12}}\ ,
\end{equation}
and for all studied $Z$
\[
    A^{GS}_1\ > \ A^{GS}_2\ .
\]
The rate of convergence of variational integrals is defined by $A^{GS}_2$. Interestingly, for the ground state of negative hydrogen ion H${}^-$ the similar consideration was presented a long ago by S.~Chandrasekhar \cite{Chandrasekhar:1944}. In this article it was emphasized the importance to study the behavior of the wavefunctions at large distances in order to get the accurate results.

In Fig.\ref{A2} the dependence on $A^{GS}_2, A^{ES}_1$ {\it versus} $Z$, found numerically, is shown. Systematically, $A^{GS}_2\ > \ A^{ES}_1$. With high accuracy both curves self-intersect and simultaneously vanish(!) at $Z=Z_B \sim 0.905$\,, which corresponds to the so-called second critical charge $Z_B$, see \cite{MPLA:2019,AOP:2019} and references therein, where both functions loose their square-integrability. It is an indication that at $Z=Z_B$ there is the second-order branch point in energy in $Z$-plane with a meaning of crossing of the ground state energy with the first excited state energy as was predicted in \cite{MPLA:2019}. It leads to appearance of the Puiseux expansion of the energy at $Z=Z_B$ \cite{Canadian:2016}. It is natural to assume that namely this branch point defines the radius of convergence of $1/Z$ expansion for both states. With high accuracy both $A^{GS}_2, A^{ES}_1$ are interpolated by
\[
      b_{1/2} (Z-Z_B)^{1/2} + b_{1} (Z-Z_B)
\]
where $b_{1/2}$ is small, while $b_{1}=1$ for $A^{GS}_2$ and $b_{1}=1/2$ for $A^{ES}_1$ in agreement with shell model, see (\ref{GS-rates-exact}), (\ref{ES-rates-exact}).

It is interesting that for $Z=1$ both $A^{GS}_2, A^{ES}_1$ take positive values, $0.117222\,(a.u.)^{-1}$\ \footnote{In \cite{Chandrasekhar:1944} it was found that $A^{GS}_2$ takes much larger value $\sim 0.47758$ even though its variational energy, being slightly worse, differs in the 3rd figure.}
and $0.047625\,(a.u.)^{-1}$, respectively, which indicate to square-integrability of functions $\Psi_{G,F}$, respectively \footnote{At $Z=1$ the potential
of interaction of neutral core (hydrogen atom) with distant electron is attractive, it behaves like $-1/r^4$ and generates van-der-Waals minimum, for discussion see e.g. \cite{Andersen:2004}}. Corresponding energies are $E^{GS}=-0.52725$\,a.u. , which is very close to the well-known exact value $-0.52775$\,a.u. (they differ in the 4th d.d.), for the ground state of negative ion of hydrogen, H$^-$, while $E^{ES}=-0.49998$\,a.u. for the 1st excited state, respectively. For the latter state the energy is very close to threshold $E_{threshold}=-0.5$\,a.u. being above of it: it is the level embedded to continuum within the accuracy which the function $\Psi_G$ provides \footnote{Mathematicians proved the existence of finite number of bound states for H$^-$ without specifying how many, see \cite{Yafaev:1974}.}. The interesting question what would happen if more accurate function than $\Psi_G$ is taken as the trial function, will this level become the bound state with energy below threshold - it is not clear to the present authors, it might be a subject of separate study.

It is well know that there exists the critical charge $Z_{cr}=0.911028\ldots$, for which the exact ground state energy $E^{GS}=-0.41496\ldots$\,a.u., calculated variationally in \cite{Estienne:2014} and confirmed in Lagrange Mesh Method in \cite{OT:2015} with high accuracy, coincides with corresponding threshold of the continuous spectra. In both calculations \cite{Estienne:2014} and \cite{OT:2015} the ground state function at $Z=Z_{cr}$ is square-integrable in agreement with the general theory, see e.g. \cite{Hoffmann:1983}. In present calculation both the ground state function $\Psi_G$, see Part I, and the spin-singlet first excited state $\Psi_F$ function remain square-integrable, since both $A^{GS}_2, A^{ES}_1$ take positive values at $Z=Z_{cr}$. The excited state at $Z=Z_{cr}$ continues to correspond to the level embedded to continuum. In both states the system is of finite size. In further decrease of $Z < Z_{cr}$ both wavefunctions remain square-integrable, $A^{GS}_2, A^{ES}_1 > 0$  - the size of the system is finite for both states in agreement with prediction by Stillinger-Stillinger \cite{Stillinger:1974}, but contrary to that was stated in \cite{Lieb:2014}. At $Z=Z_B$ the square-integrability of both functions is lost simultaneously and the size of the system gets infinite for both spin-singlet
states.

As a conclusion we state that it was constructed successfully the accurate trial functions for two spin-singlet states $(1s\,1s\,1{}^1S)$ ($\Psi_F$) and $(1s\,2s\,2{}^1S)$ ($\Psi_G$) of Helium-like sequence. These functions are orthogonal by construction. One can calculate matrix  elements of the Hamiltonian,
\[
   H_{FF}=<\Psi_F|H|\Psi_F>\ =\ E_F,\ H_{GF}=<\Psi_G|H|\Psi_F>=H_{FG}\ ,\ H_{GG}=<\Psi_G|H|\Psi_G>=E_G \ ,
\] 
and calculate the eigenvalues of the corresponding 2 x 2 symmetric matrix. It is evident that
non-diagonal matrix element $H_{GF}$ is very small in comparison with diagonal ones, the eigenvalues will differ from variational energies slightly. This procedure leads to the improvement of the variational energies.

\begin{figure}
\caption{
\label{A2}
   He-like sequence, $(1s2s \, 2 {}^1 S)$ and $(1s1s \, 1 {}^1 S)$ states:\\
  (a) the parameter $A^{(ES)}_1$ {\it vs } $Z$ for the functions  $\Psi_{HK}$ (\ref{He-infty-psi-2}) and $\Psi_{G}$ (\ref{He-infty-psi-3})
  shown (lower curves), which ``measures" their square-integrability , see text. For $Z \leq 3/2$ the dot-dashed part of the curve and dotted curve are the extrapolation, see embedded subfigure, \\
  (b) for the ground state $(1s1s \, 1 {}^1 S)$ the parameter $A^{(GS)}_2$ {\it vs } $Z$ for the function $\Psi_{F}$, see Part I, Eq.(23), is shown (upper curve), it almost coincides with curve $A^{(GS-HK)}_2$ (not shown) with $Z^{(GS-HK)}_B \sim 0.887$.\\
  (c) With high accuracy the curves $A^{(ES-G)}_1$ and $A^{(GS-F)}_2$ intersect at $Z=Z^{(ES-G)}_B=Z^{(GS-F)}_B \sim 0.905$, where square-integrability is lost for both states simultaneously. Dash-dotted curve corresponds to Hylleraas-Kinoshita function (\ref{He-infty-psi-2}) with $Z^{(ES-HK)}_B \sim 0.875$.
}
  \includegraphics[scale=0.5,angle=-90]{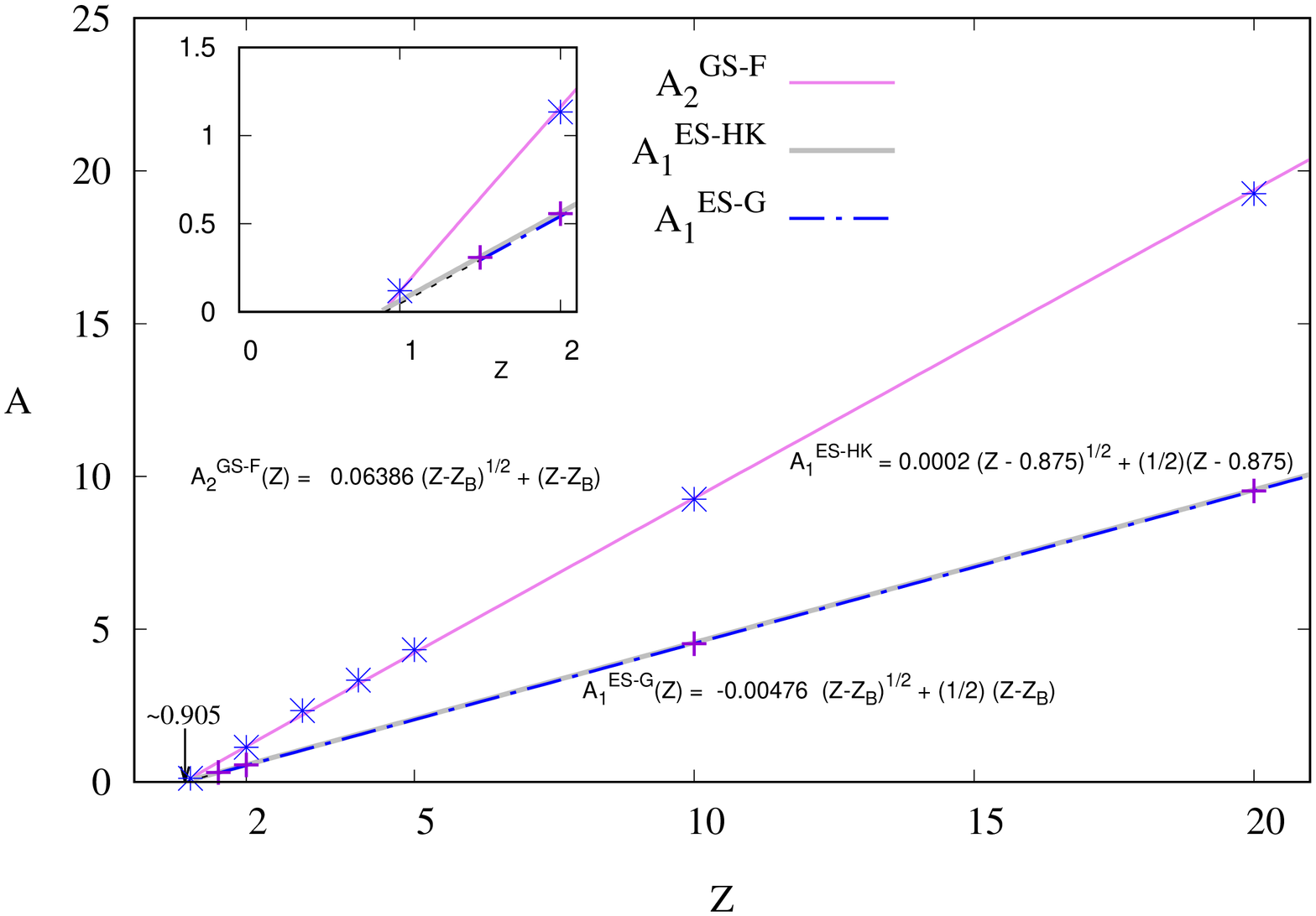}
\end{figure}

\newpage

\section{Spin-triplet state $1{}^3S$ (ortho-Helium sequence): generalizing Hylleraas function}

\subsection{Generalities, exact solution}

Orbital functions for spin-triplet state at zero total angular momentum depend on relative distances $(r_1,r_2,r_{12})$ between bodies and they are antisymmetric,
\[
   \Psi^{(-)} (r_1,r_2,r_{12})\ =\ -\,\Psi^{(-)} (r_2,r_1,r_{12})\ ,
\]
hence, it might be written as
\begin{equation}
\label{psi-}
   \Psi^{(-)} (r_1,r_2,r_{12})\ =\ (1 - P_{12})\, \phi (r_1,r_2,r_{12})\ =\
   \phi (r_1,r_2,r_{12}) - \phi (r_2,r_1,r_{12})\ .
\end{equation}
It implies the existence of nodal surface
\[
   r_1\ =\ r_2\ .
\]
It suggests to employ the representation
\[
  \Psi^{(-)}(r_1,r_2,r_{12})\ =\ (r_1\ -\ r_2)\,\psi(r_1,r_2,r_{12})\ ,
\]
where $\psi$ is symmetric, $\psi (r_1,r_2,r_{12})\ =\ \psi (r_2,r_1,r_{12})$.
Permutationally-symmetric function $\psi$ is the eigenfunction of the gauge rotated Hamiltonian,
\begin{equation}
\label{H-}
\hat{\mathcal{H}}\ =\ (r_1-r_2)^{-1}\,\hat{H}\,(r_1-r_2)\ =\
\end{equation}
\begin{gather}
       -\frac{1}{2}
\left[
  \frac{\pa^2}{\pa r_1^2}\ +\ \frac{\pa^2}{\pa r_2^2}\ +\ 2\frac{\pa^2}{\pa r_{12}^2}\ +\ 2\left(\frac{1}{r_1-r_2}+ \frac{1}{r_1}\right)\frac{\pa}{\pa r_1}
  \ +\ 2\left(\frac{1}{r_2-r_1} + \frac{1}{r_2}\right)\frac{\pa}{\pa r_2}
\right.
\non
\\
\left.
  +\ \left[\frac{r_1^2+6r_1r_2+r_2^2-r_{12}^2}{r_1 r_2 r_{12}}\right]\frac{\pa}{\pa r_{12}}
   \ +\
   \left[\frac{r_1^2-r_2^2+r_{12}^2}{r_1 r_{12}}\right]\frac{\pa^2}{\pa_{r_1}\pa_{r_{12}}}
   \ +\
   \left[\frac{r_2^2-r_1^2+r_{12}^2}{r_2 r_{12}}\right]\frac{\pa^2}{\pa_{r_2}\pa_{r_{12}}}
\right]
\non\\
   +\ \left[\frac{-Z}{r_1}\ +\ \frac{-Z}{r_2}\ +\ \frac{\delta}{r_{12}}+ \frac{1}{r_1r_2}\right]
   \ ,
\non
\end{gather}
where $\hat{H}$ is the original (reduced) Hamiltonian acting in $(r_1,r_2,r_{12})$ space, see \cite{GAM:1987} and \cite{Part-1:2020}; here the parameter $\de=1$ - it is introduced for the sake of convenience. Eventually, the eigenvalue problem
\begin{equation}
\label{SE}
       \hat{\mathcal{H}}\psi\ =\ E\psi\ ,\quad \psi (r_1, r_2, r_{12}) \in L^2({\bf R}^3_+)
       \ ,
\end{equation}
occurs where the condition $\psi(0,0,0) = 0$ should be imposed.

At $Z \rar \infty$ (equivalently, $\de=0$) for the lowest energy spin-triplet state $(1s2s\,1{}^3S)$, there exists the exact eigenfunction in the form of anti-symmetrized product of $(2s1s)$ orbitals,
\begin{equation}
\label{He-infty-psi-}
  \Psi^{(-)}_0\ =\ \frac{1}{2}(1 - P_{12}) (1 - a Z r_1)\,e^{-\al Z r_1 - \beta Z r_2}\ \sim
  (2s_11s_2) - (1s_12s_2)\ ,
\end{equation}
where $a=1/2, \al=1/2, \beta=1$; they are related as
\begin{equation}
\label{constraint}
   a\ =\ -\al+\beta\ ,
\end{equation}
with eigenvalue
\begin{equation}
\label{He-infty-E-}
  E_0\ =\ -  \frac{5 Z^2}{8} \ ,
\end{equation}
cf. (\ref{He-infty-psi-1}),  (\ref{He-infty-E-1}).

Making the analysis of the Schr\"odinger equation for the Hamiltonian (\ref{H-}) it can be shown that at small distances a solution admits the analogue of the Fock expansion for the ground state
\[
  \Psi(r_1,r_2,r_{12})\ =\ (r_1\ -\ r_2)
\]
\begin{equation}
\label{Fock-}
  \left(K\ +\ (r_1\ +\ r_2)\ +\ D_0\, r_{12}\ -\ A\,\left(r_1^2+r_2^2\right)-
  B\, r_1 r_2\ +\ C\,(r_1+r_2)r_{12}\ +\ D\,r_{12}^2\ +\ \ldots \right)
\end{equation}
with
\[
   A\ = \frac{2Z}{3}\quad ,\quad  B\ =\ \frac{5Z}{3}\quad ,\quad C\ =\ \frac{\de}{4}\quad ,\quad K\ =\ D_0\ =\ D\ =\ 0\ ,
\]
thus, the constant term is expansion of the second factor is absent as well as the terms $r_{12}, r_{12}^2$. We call the cusp parameters the following expressions
\begin{equation}
\label{cusps-3}
   C_{Z,e}^{(1)}= (-A + B) = Z\quad ,\quad  C_{Z,e}^{(2)} = (4A-B) = Z\quad ,
   \quad C_{e,e}\ =\ 4 C\ =\ \de\ (=\ 1)\ ,
\end{equation}
having the meaning of residues at Coulomb singularities of the potential.
Alternatively, the electron-nuclear cusps can be calculated through the ratio of expectation values
\begin{equation}
\label{cusp-Ze}
 {\tilde C}^{(i)}_{Z,e}\ \equiv\ -\ \frac{\langle \de({\bf r}_i)\frac{\pa}{\pa r_i} \rangle}{\langle
   \de({\bf r}_i) \rangle}\ ,\ i= 1,2\qquad ,\qquad
\end{equation}
see \cite{CS:1969} and also \cite{Part-1:2020}. For the exact eigenfunction $ {\tilde C}^{(1,2)}_{Z,e}=Z$.

\subsection{Compact trial functions for spin-triplet state}

A straightforward way to construct trial functions for spin triplet state is by making a generalization of the procedure used for spin singlet state
\begin{equation}
\label{first}
  \Psi_a^{(-)}\ =\ (1 - P_{12})\, \Psi_0 (r_1,r_2,r_{12})\ =\ \Psi_0 (r_1,r_2,r_{12}) - \Psi_0 (r_2,r_1,r_{12})\ .
\end{equation}
Imposing the condition that $\Psi_a^{(-)}$ should reproduce at $Z \rar \infty$  the exact solution (\ref{He-infty-psi-}), one can construct the negative parity analogue of Hylleraas function,
\begin{equation}
\label{HeZ-}
  \Psi_{H}^{(-)}\ =\ \frac{1}{2}(1 - P_{12})\,(1 - a Z r_1) e^{ - \al Z r_1 - \beta Z r_2 + \gamma r_{12}}
  \ ,
\end{equation}
This function is characterized by four free parameters with one constraint (\ref{constraint}),
$a\ =\ -\al+\beta$, due to boundary condition for (\ref{SE}), or equivalently, $K=0$ in (\ref{Fock-}).
Extra conditions $D_0\, =\, D\, =\, 0$ from (\ref{Fock-})
- the absence of the terms $r_{12}, r_{12}^2$ - are fulfilled automatically.
Cusp parameters are easily found,
\begin{equation}
\label{cusp-hylleraas}
   C_{Z,e}^{(1)}\ =\ \beta\, Z\quad ,\quad C_{Z,e}^{(2)}\ =\ 2\al \,Z \quad ,
   \quad C_{e,e}\ =\ \gamma\ .
\end{equation}
The variational calculations can be carried out analytically, the results are presented in Table II.

There is another trial function, which takes into account the effect of screening of electron-electron interaction due to presence of charge $Z$ and which seems essential,
it appears as the generalization of the Hylleraas function
(\ref{HeZ-}),
\begin{equation}
\label{psinew}
 \Psi_F^{(-)}\ =\ \frac{1}{2}\,(1 - P_{12})\,\bigg[
 \left(1 + (\al - \beta) Z r_1\right)
 \,e^{- \al Z r_1\, -\, \beta Z r_2\, +\, \gamma r_{12}\, \frac{(1 + c r_{12})}{(1 + d r_{12})}}\bigg]\ ,
\end{equation}
where $\al, \beta, \gamma, c, d$ are 5 free parameters; $K=0$ and conditions $D_0\, =\, D\, =\, 0$ from (\ref{Fock-}) in the expansion of $\Psi_F^{(-)}$ at small distances continue to be fulfilled.
The cusp parameters are the same as in (\ref{cusp-hylleraas}). The results of variational calculations are presented in Table II.

The function (\ref{psinew}) can be further generalized to
\begin{equation}
\label{psinew-final}
 \Psi_G^{(-)}\ =\ \frac{1}{2}\,(1 - P_{12})\,\bigg[
 \left(1 + (\al - \beta )  Z r_1  -a (r_1 +   r_2)\right)
 \,e^{- \al Z r_1\,\frac{(1 + c_1 r_{1})}{(1 + d_1 r_{1})}\,-\,\beta Z r_2\,
  +\, \gamma r_{12}\,\frac{(1 + c_{12} r_{12})}{(1 + d_{12} r_{12})}}\bigg]\ ,
\end{equation}
where the screening of interaction of $(2s)$ electron, marked by $1$, with charge $Z$ is taken into account as well as the screening of interelectron interaction.
This function is going to be used in final calculations, see Table II. This function contains 8 free parameters; the cusp parameters are,
{\small
\begin{align}
\label{cusp-psig}
C_{Z,e}^{(1)} &= \, \frac{
\,   \left( (\al\, - \beta ) \, \beta\,Z - (\al\,+ \beta)\,{a}\right)\,Z
+ 2\, \tilde\al\,\de_1    \left( a+\beta\,Z \right)
}
 {      (\al - \beta)\,Z - 2\, a  \,  + 2 \tilde\al \, \de_1 }\ ,
 \non \\[5pt]
C_{Z,e}^{(2)} &= \, 2 \frac{
\,   \left(
({\alpha}-\,\beta)\,\al\,Z -( \al\,+\beta)\,{a}
 \right) {Z}
+2\, \tilde\alpha\, \delta_1    \left( {a} + \beta\,Z \right)
+3\,    {d_1}\,   \tilde\al\, \delta_1
}
 {     (\al - \beta)\,Z - 2\, a  \,  + 2  \tilde\al \, \de_1 }\ ,
 \\[5pt]
C_{e,e} &= \gamma\ ,
\non
\end{align}
where $ \delta_1 = c_1 -d_1\,$ and $\tilde\alpha = \alpha/(\alpha-\beta)$. Further generalization does not look necessary. It does lead to improvement of energy $E_{var}$ obtained with $\Psi^{(-)}_{G}$ (see Table II, column 5) in 5th decimal digit, which is beyond the scope of the present paper.}

Imposing three constraints on parameters in (\ref{psinew-final}), see (\ref{cusp-psig}), in particular, choosing $\gamma=1/4$, one can reproduce cusp values (\ref{cusps-3}) exactly.
The energies obtained with such constrained parameters (after making minimization with respect to five free parameters) for the function (\ref{psinew-final}) are presented in Table II, column 4. The remaining free parameters can be easily fitted by the following functions
\begin{align}
   \al Z   &\ =\  -0.16755  + 0.37339 Z + 0.00762  Z^2 \ , \non \\
   \beta Z &\ =\   0.005394 + 1.00163 Z - 0.000086 Z^2 \ , \\
    c_{12} &\ =\  -0.06847  + 0.07471 Z - 0.00537  Z^2 \ , \non \\
    d_{12} &\ =\  -0.07878  + 0.47309 Z - 0.02018  Z^2 \ , \non \\
    c_1    &\ =\   1.11157  + 0.43103 Z - 0.02155  Z^2 \ . \non
\end{align}
These fitted parameters allow us to obtain expectation values for energies for any value $Z \leq 20$ with sufficiently high accuracy, see Table II. In particular, for $Z=15$ the fitted parameters (31) lead to the expectation value for energy $E=-137.825$\,a.u., while the Majorana formula (\ref{Majorana-}), see below, gives $E= -137.854$\,a.u.

Interestingly, for $Z=2$ all trial functions used allow us to reproduce 4 s.d. in energy, while $\Psi_G^{(-)}$ (unconstrained) result is in agreement with benchmark results \cite{Korobov:2018} in 5 s.d. as well as \cite{Accad:1971} and \cite{Braun:1993}. Note that for $Z=10,20$ functions $\Psi_F^{(-)}$ and $\Psi_G^{(-)}$ in spite of their simplicity lead to the lower energies in comparison with old benchmark calculations \cite{Drake:1988}. For $Z=10$ they differ from one in \cite{Pachucki:2010} in 6th s.d. and establish benchmark for $Z=20$.

\begin{table}[htb]
\begin{center}
        \caption{Comparison of the lowest energy spin-triplet state $(1s2s\,1{}^3S)$ for Helium-like atomic ions at $Z=3/2,2,10,20$ calculated variationally using the Hylleraas function  $\Psi^{(-)}_H$ (\ref{HeZ-}), with the function $\Psi^{(-)}_{F}$ (\ref{psinew}) and with function
        $\Psi^{(-)}_{G}$ (\ref{psinew-final}) (constrained, where it is imposed: cusp values $C_{Z,e}^{(1)}=C_{Z,e}^{(2)}=Z, C_{e,e}=1$ reproduced {\it exactly}), and (un-constrained, where all 8 parameters free).
        The energy $E_{var}$ in a.u., the nuclear-electron cusps $C_{Z,e}$ and electron-electron  cusp $C_{e,e}$ shown. Last column shows energies from \cite{Korobov:2018}${}^a$, \cite{Drake:1988}${}^b$, \cite{Pachucki:2010}${}^c$}
\label{H-vs-F}
{\setlength{\tabcolsep}{0.1cm} \renewcommand{\arraystretch}{1.5}
 \resizebox{0.9\textwidth}{!}{%
                \begin{tabular}{c|ccc|ccc|ccc|ccc|c|cc}
                        \hline
                        \multirow{2}{*}{$Z$} &
                        \multicolumn{3}{c|}{ $\Psi^{(-)}_H$} &
                        \multicolumn{3}{c|}{ $\Psi^{(-)}_F$} &
                        \multicolumn{3}{c|}{ $\Psi^{(-)}_{G}$ (constrained)}    &
                        \multicolumn{3}{c|}{ $\Psi^{(-)}_{G}$ (un-constrained)} &
                        \multirow{2}{*}{$E_{\rm refs}$} &
                        \multirow{2}{*}{$E_{\rm Majorana}$}\\
              & $E_{var}$   &    $C^{(1)}_{Z,e}/C^{(2)}_{Z,e}$\   &\    $C_{e,e}/4$\   &  $E_{var}$ & $C^{(1)}_{Z,e}/C^{(2)}_{Z,e}$ \ & \
               $C_{e,e}/4$ \ & $E_{var}$ & $C^{(1)}_{Z,e}/C^{(2)}_{Z,e}$\ &\ $C_{e,e}/4$ & $E_{var}$ & $C^{(1)}_{Z,e}/C^{(2)}_{Z,e}$\ &\ $C_{e,e}/4$ &
               \\[4pt]
  \hline\hline
 \rule{0pt}{4ex}
 3/2       & -1.1773   & 1.504  & 0.026 & -1.1774   & 1.504  & 0.25  & -1.1774    & 1.500  & 0.250 &  -1.17758  & 1.517  & 0.206 &                     &   -1.17482        \\
           &           & 0.803  &       &           & 0.802  &       &            & 1.500  &       &            & 0.982  &       &                     &                   \\[4pt]
 2         & -2.1749   & 2.007  & 0.035 & -2.1750   &  2.006 & 0.25  & -2.1750    & 2.000  & 0.250 & -2.17515   & 2.019  & 0.214 & -2.17523$\,{}^a$    & -2.1745       \\
           &           &  1.351 &       &           &  1.349 &       &            & 2.000  &       &            & 1.482  &       &    -2.17521$\,{}^b$ &               \\
           &           &        &       &           &        &       &            &        &       &            &        &       &    -2.17523$\,{}^c$ &               \\[4pt]
 \mbox{10} & -60.6681  & 10.013 & 0.053 & -60.6684  & 10.011 & 0.250 & -60.66847  & 10.000 & 0.250 & -60.66852  & 10.036 & 0.225 & -60.66836$\,{}^b$   & -60.6694       \\
           &           &  9.415 &       &           &  9.411 &       &            & 10.000 &       &            & 9.538  &       & -60.66864$\,{}^c$   &                \\[4pt]
\mbox{20}  & -246.2886 & 20.014 & 0.054 & -246.2888 & 20.012 & 0.250 & -246.28838 & 20.000 & 0.250 & -246.28896 & 20.036 & 0.231 & -246.28847$\,{}^b$  &-246.28796       \\
           &           & 19.420 &       &           & 19.412 &       &            & 20.000 &       &            & 19.501 &       &                     &                 \\[4pt]
\hline
\end{tabular}}}
\end{center}
\end{table}

\begin{table}[htp]
\caption{Expectation values for the excited state $(1s2s\,1{}^3S)$ using the function $\Psi_G^{(-)}$ constrained/unconstrained.
Results by Accad-Pekeris-Schiff  \cite{Accad:1971}, Braun-Schweizer-Herold \cite{Braun:1993}
and Liverts-Barnea \cite{Liverts} ${}^{(\star)}$ included for comparison.}
\begin{center}
{\setlength{\tabcolsep}{0.1cm} \renewcommand{\arraystretch}{0.5}
 \resizebox{0.9\textwidth}{!}{%
\begin{tabular}{|c|llllccccccc|}\hline
 $Z$  &  $E$   & $C_{Z,e}^{(1)}$ &  $C_{Z,e}^{(2)}$ &  $C_{e,e}/4$ & $\langle \delta({\mathbf r_1})\rangle$ & $\langle \delta({\mathbf r_{12}})\rangle$
      & $\langle r_{1}\rangle$ & $\langle  r_{1}^2 \rangle$  & $\langle  r_{12} \rangle$
      & $\langle r_{12}^2 \rangle$ & source\\
\hline
      & 	& 	& 	&         &          &   &    &    &   &   \\[3pt]
 3/2  & -1.17739 & 1.5   & 1.5   & 0.25  & 0.54779 & 0 & 4.07023 & 30.48296 & 7.24355 & 61.24372 &
\\[3pt]
      & -1.17758 & 1.517 & 0.982 & 0.206 & 0.54595 & 0 & 4.16499 & 32.52315 & 7.43475 & 65.31719 &
\\[15pt]
  2   & -2.17501 &  2     & 2      & 0.25   & 1.32437 & 0 & 2.53536 & 11.23822 & 4.41675 &
      22.60109 &
\\[3pt]
      & -2.17515 & 2.019 & 1.482 & 0.214 & 1.32265 & 0 & 2.55238 & 11.49963 & 4.45021 &  23.10662 &
\\[3pt]
      & -2.17523 &        &        &        & 1.32036 & 0 & 2.55046 & 11.46432 & 4.44754 & 23.04620 &  \cite{Accad:1971}\\[3pt]
	                &-2.17523                &        &        &        &1.32035&0   &2.55047               &                 &4.44753             &                 & \cite{Liverts}\\[3pt]
      & -2.17522 &        &        &        & 1.32028 &   & 2.55047 & 11.46438 &         &                  & \cite{Braun:1993}
\\[15pt]
 10   & -60.66847 & 10    &  10    & 0.25   & 175.8732 & 0 & 0.39261 & 0.24991 & 0.66026 &
      0.50081 &
\\[3pt]
      & -60.66852 & 10.036 &   9.538   & 0.225 & 175.8991 & 0 & 0.39277  & 0.25034 &
      0.66058   & 0.50161   &
\\[3pt]
      & -60.66835 &         &            &        & 175.7860 & 0 & 0.39276   & 0.25027 &
      0.66061   & 0.50155   &  \cite{Accad:1971}
\\[3pt]
				      &-60.66865	&      &            &      &175.7857& 0&        0.39276&   &0.66061    &    & \cite{Liverts}\\[15pt]
 20   & -246.28838 & 20     &  20        &  0.25  & 1420.1148 &  0 & 0.19178 & 0.059305 &
      0.321414  & 0.11872  &
\\[3pt]
      & -246.28896 & 20.036 & 19.501   & 0.231 & 1419.6493 &  0 & 0.19178 & 0.059248 &  0.321362  & 0.118616  &
\\[3pt]
					&-246.28908&               &              &           &1419.1334  &  0  &0.19177             &                &0.32137                    &   & \cite{Liverts}
					\\[10pt]
\hline
\end{tabular}}}
		\begin{tablenotes}
			\small
            \item $^{\star}$ calculated using the code provided in \cite{Liverts}
		\end{tablenotes}
\end{center}
\label{table4}
\end{table}%

It can be shown that the Majorana formula holds for this state
\begin{equation}
\label{Majorana-}
 E^{(1^3S)}\ =\  -\,\frac{5 Z^2}{8}\ +\ 0.188141\, Z\ -\ 0.0507843\ .
\end{equation}
It provides accuracy 4 s.d. for $2 \leq Z \leq 10$ and 5 s.d. for $Z \geq 10$! In general, the energies of the excited states $(1s2s\,2{}^1S)$ and $(1s2s\,1{}^3S)$ are close to each other, while at $Z \rar \infty$ these states become degenerate, see (\ref{He-infty-E-1}) and (\ref{He-infty-E-}), respectively.

In Table \ref{table4} six expectation values for the excited state $(1s 2s 1^3S)$ using the function $\Psi_G^{(-)}$ are presented in the two cases of the constrained function which reproduces the cusp parameters exactly, and unconstrained one. There exists the agreement with results by Accad-Pekeris-Schiff \cite{Accad:1971}, Braun-Schweizer-Herold \cite{Braun:1993} and \cite{Liverts}, wherever possible, in 2-3-4 s.d.

A simple interpolation of the energies obtained with the un-constrained function $\Psi_G^{(-)}$ (see Table \ref{H-vs-F}) shows that the critical charge for which the system has a vanishing ionization energy is $Z=Z_{cr} \simeq 1.085$. At this point the function $\Psi_G^{(-)}$ remains square-integrable.

\subsection{Square-integrability of compact functions}

As was stated in Section I.D if keeping the position of one electron fixed the wavefunction decays exponentially at large values of the position of another electron, at least for $Z > 1$. It can be seen explicitly if we take the first term in $\Psi^{(-)}(r_1,r_2,r_{12})$ in (\ref{psi-}).  Similar relations as (\ref{phirinfinity})  are obtained for $ \phi|_{r_2(r_1) fixed}$.
For the exact solution (\ref{He-infty-psi-}) the corresponding parameters $A_{1,2}$ in  (\ref{phirinfinity})  are,
\[
     A^{(-)}_1\ =\ \frac{1}{2} Z\ ,\ A^{(-)}_2\ =\ Z\ ,
\]
they correspond to the Coulomb charges on $2s(1s)$ orbitals.
It implies that at $Z=Z^{(-)}_B=0$ square-integrability of (\ref{He-infty-psi-}) is lost.

Taking the first term in $\Phi^{(-)}_G(r_1,r_2,r_{12})$ in (\ref{psinew-final}) we arrive at the parameters for the lowest spin-triplet excited state (ES3),
\begin{equation}
\label{AES-3}
     A^{ES3}_1\ =\ \alpha Z \frac{c_1}{d_1} - \gamma \frac{c_{12}}{d_{12}}\ ,\
     A^{ES3}_2\ =\ \beta Z  - \gamma \frac{c_{12}}{d_{12}}\ .
\end{equation}
It turns out that in concrete calculations, for all studied $Z$,
\[
    A^{ES3}_1\ < \ A^{ES3}_2\ ,
\]
for both functions $\Psi_{H}^{(-)}$ and $\Psi_{G}^{(-)}$ as the appearance of clusterization effect. In Fig.\ref{Atriplet} the behavior of $A^{ES3}_1$ {\it vs} $Z$ for functions $\Psi_{H}^{(-)}$ and $\Psi_{G}^{(-)}$ is shown. With high accuracy for both functions it is a straight line with slope $1/2$. At $Z=Z^{(ES3)}_{B} \simeq 0.90 \,(0.76)$ it crosses the horizontal line where the function $\Psi_G^{(-)}$ $(\Psi_{H}^{(-)})$ becomes non-square-integrable. It is natural to assume that the $Z$-complex plane of energy has the square-root branch point at $Z=Z^{(ES3)}_{B}$, where the energy level $1{}^3S$ crosses with the energy level $2{}^3S$. It seems likely, that in this point the Puiseux expansion in $(Z-Z^{(ES3)}_{B})$ can be built, cf. \cite{AOP:2019} and references therein. It will be done elsewhere.
\begin{figure}
\caption{
\label{Atriplet}
 He-like sequence, state $(1s2s\,1{}^3S)$:\\ the parameter $A_1$ {\it vs } $Z$, see text:
 (\ref{AES-3}) for the function $\Psi^{(-)}_{G}$ (\ref{psinew-final}) (with un-constrained parameters)
 found numerically (marked by crosses) and fitted (solid line) with $Z^{(ES3)}_B \sim 0.902$ compared
 with $A_1=(\al Z - \gamma)$ for (\ref{HeZ-}) - dashed line, which predicts $Z_B \sim 0.760$;
 $A^{(H,G)}_1$ ``measures" the square-integrability of functions
 (\ref{psinew-final}) and (\ref{HeZ-}), respectively.
}
  \includegraphics[scale=0.5,angle=-90]{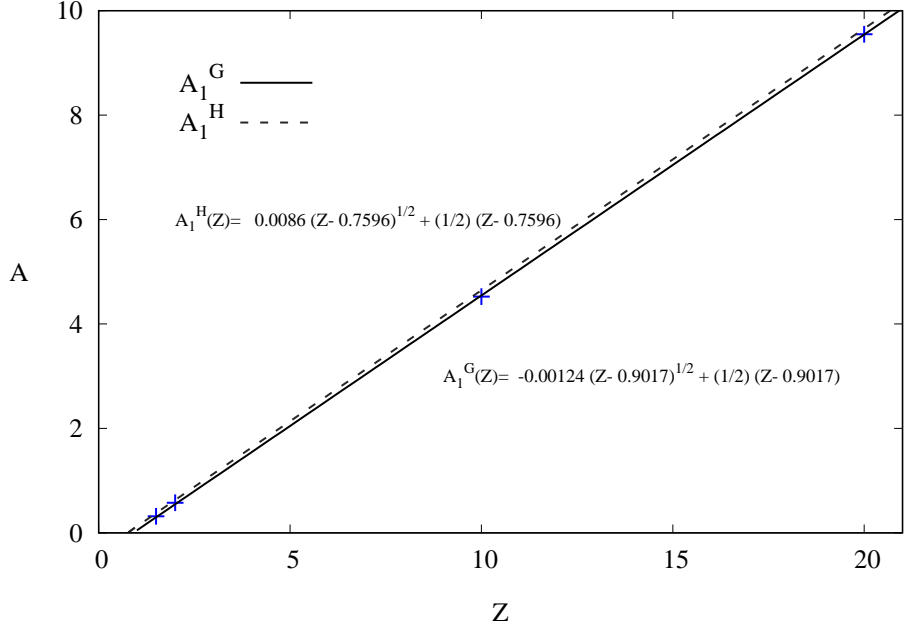}
\end{figure}

\section*{Conclusions}

In this article the few-parametric ultra-compact variational trial functions are constructed for spin-singlet $(1s 2s 2^1 S)$ and spin-triplet $(1s 2s 1^3 S)$ states for He-like
iso-electronic sequence with $Z \in [2,20]$. In the limit $Z \rar \infty$ these functions
become the exact eigenfunctions for the problem of two non-interactive Hydrogen atoms.
At $Z \in [2,20]$ they allow us to reproduce the energies with 4-5 s.d. and some expectation values with 2-3 s.d. while the cusp parameters are reproduced exactly. We assume that all obtained figures can not be changed by mass, relativistic and QED corrections. With high accuracy these figures are reproduced by the Majorana formula - the second degree polynomial in $Z$.

In general, the atomic ions with $Z > 2$ are poorly studied and many of our results are presented for the first time. It turned out that the surprising observation by Bressanini-Reynolds about a special form of nodal surface for $2^1 S$ state made for Helium atom can be extended to the atomic ions with $Z > 2$. It is crucial for Monte-Carlo studies of the excited states of highly charged ions. Methodology which we presented can be easily extended for building the ultra-compact functions for other excited states of He-like sequence. It will be done elsewhere.

In the Part III \cite{Part-3:2021} the spin-quartet state $1\,{}^40^+$ of Li-like sequence will be studied using ultra-compact trial functions for $Z \leq 20$.

\section*{Acknowledgments}

\noindent
A.V.T. thanks PASPA-UNAM for support during his sabbatical leave in the last stage of the work. 
J.C.~del\,V. is supported by CONACyT PhD Grant No.570617 (Mexico) in the early stage of the work and by postdoctoral grant via DGAPA grant IN113819 (Mexico) in the late stage of the work. This work is partially supported by CONACyT grant A1-S-17364 and also DGAPA grant IN113819 (Mexico). D.J.N. is supported in part by PRODEP project 42027 UV-CA-320 (Mexico).

\end{document}